\documentclass[a4paper]{jpconf}
\usepackage[english]{babel}
\usepackage[utf8]{inputenc}
\usepackage[T1]{fontenc}
\usepackage{amsmath}
\usepackage{amssymb}
\usepackage{enumitem}
\usepackage[separate-uncertainty=true,
 range-units = single,
 range-phrase = --
]{siunitx}

\usepackage{graphicx}

\begin{document}
\title{Status of the Lunar Detection Mode for Cosmic Particles of LOFAR }

\author{T.~Winchen$^{1}$ A.~Bonardi$^{2}$, S.~Buitink$^{1}$, A.~Corstanje$^{2}$, H.~Falcke$^{2,3,4}$, B.~M.~Hare$^{5}$, J.~R.~Hörandel$^{2,3}$, P.~Mitra$^{1}$, K.~Mulrey$^{1}$, A.~Nelles$^{6}$, J.~P.~Rachen$^{2}$, L.~Rossetto$^{2}$, P.~Schellart$^{2,7}$, O.~Scholten$^{5,8}$, S.~ter~Veen$^{2,4}$, S.~Thoudam$^{2,9}$, T.~N.~G.~Trinh$^{5}$}

\address{$^1$ Astrophysical Institute, Vrije Universiteit Brussel, Pleinlaan 2, 1050 Brussels, Belgium,}
\address{$^2$ Department of Astrophysics/IMAPP, Radboud University, P.O. Box 9010, 6500 GL Nijmegen, The Netherlands,}
\address{$^3$ NIKHEF, Science Park Amsterdam, 1098 XG Amsterdam, The Netherlands,}
\address{$^4$ Netherlands Institute of Radio Astronomy (ASTRON), Postbus 2, 7990 AA Dwingeloo, The Netherlands,}
\address{$^5$ KVI-CART, University Groningen, P.O. Box 72, 9700 AB Groningen,\\}
\address{$^6$ Institut für Physik, Humboldt-Universität zu Berlin, Unter den Linden 6, 10099 Berlin, Germany,}
\address{$^7$ Department of Astrophysical Sciences, Princeton University, Princeton, NJ 08544, USA,}
\address{$^8$ Interuniversity Institute for High-Energy, Vrije Universiteit Brussel, Pleinlaan 2, 1050 Brussels, Belgium,}
\address{$^9$ Department of Physics and Electrical Engineering, Linn\'euniversitetet, 35195 V\"axj\"o, Sweden}

\ead{tobias.winchen@rwth-aachen.de}

\begin{abstract}
Cosmic particles hitting Earth's moon produce radio emission via the Askaryan
effect. If the resulting radio ns-pulse can be detected by radio telescopes,
this technique potentially increases the available collective area for ZeV
scale particles by several orders of magnitude compared to current experiments.
The LOw Frequency ARray (LOFAR) is the largest radio telescope operating in the
optimum frequency regime for this technique.  In this contribution, we report
on the status of the implementation of the lunar detection mode at LOFAR.
\end{abstract}

\section{Introduction}
The low fluxes of the particles of interest, which consequently require huge
detectors at the highest energies, is one of the major challenges in
astroparticle physics.  As a possible solution it has already been proposed in
the 1960s~\cite{Askaryan1962} to use the Moon as a particle detector by
searching for the ns radio pulses generated by a charge excess in the shower
developing in the lunar rock. Several searches for these signals have been
conducted with earth-bound radio telescopes~(see reference~\cite{Bray2016} for
a review), but so far none of these searches has been sensitive enough to
expect to detect the observed flux of charged cosmic rays. The main reason here is that
most of these previous searches used telescopes operating at GHz frequencies. While the
amplitude of the emitted pulse reaches a maximum in the GHz range, the
effective area increases towards lower frequencies as the Cherenkov cone
becomes broader and thus also inclined events become detectable. The resulting
optimum frequency band for corresponding observations is found to be above
approximately 100 MHz~\cite{Scholten2006}.

The currently largest telescope operating in this optimal frequency range is
the LOw Frequency ARray (LOFAR)~\cite{vanHaarlem2013}. It is composed of more
than 50 stations located throughout Europe with 24 stations placed in a dense
core in the Netherlands. Each of these core-stations contains 768 high band
antennas (HBA) operating in the frequency range \SIrange{110}{190}{\mega\hertz}
organized in two fields of 24 tiles of 16 antennas each. The signal of the individual stations
are filtered into
sub-bands by a polyphase filter (PPF) and combined into a beam of approximately
\SI{5}{\degree} width at \SI{120}{\mega\hertz} before it can be accessed by
users of the telescope with dedicated observation pipelines.

\section{The NuMoon Pipeline}
\begin{figure}[b]
	\includegraphics[width=\textwidth]{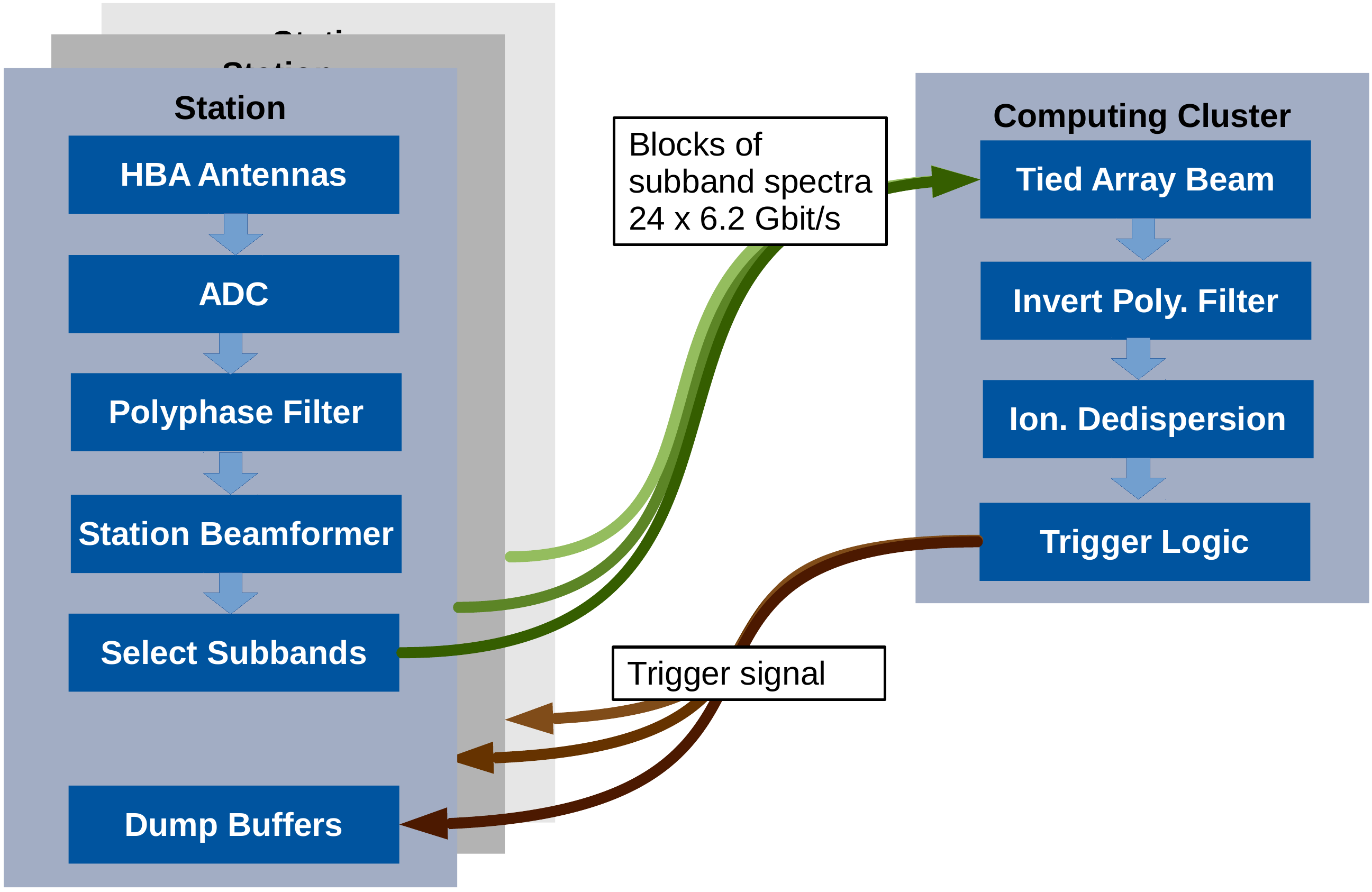}
	\caption{Data processing steps in the individual stations and central computing system for the LOFAR pipeline to observe pulses on ns timescales.}
	\label{fig:DataProcessingSteps}
\end{figure}
A pipeline to detect ns pulses with LOFAR has to perform three different tasks
to reconstruct the pulses before triggering. 
\begin{enumerate}[label=\arabic*.]
	\item Form tied array beams. To maximize the collective area, the signals of as
		many stations as possible have to be combined coherently. As this reduces
		the beam-width, several beams have to be formed to cover the surface of the
		Moon.
	\item Invert the PPF. The PPF on station level reduces the time resolution by
		a factor of 1024 from approximately ns to  approx.~$\mu$s. To recover ns time resolution, the PPF has to be inverted.
	\item Ionospheric dedispersion. By travelling through Earth's ionosphere, the
		signal is dispersed proportional to the electron content of the ionosphere. The amplitude of the pulse must consequently be reconstructed to detect the pulses.
\end{enumerate}
The corresponding data processing procedure for LOFAR is sketched in fig.~\ref{fig:DataProcessingSteps}.
In particular the inversion of the PPF is compute-intensive. A direct inversion
of the PPF produces echos of short pulses as e.g.\ shown in~\cite{Singh2012}
that limits the usability here, as we thus expect a large number of false
triggers generated from RFI by this method. Instead, we use an iterative method
that requires solving a linear system~\cite{Winchen2016a} which requires
$\mathcal{O}{(1000)}$ GFLOPs$^{-1}$ of computing power~\cite{Winchen2016b} per
beam.

The required computing resources are, unfortunately, not available on the
default LOFAR computing system. Instead, we will use the DRAGNET
cluster~\cite{DRAGNETURL}, a computing cluster build for realtime searches of
pulsars located in the same computing center next to the default LOFAR
facilities. DRAGNET is connected to the LOFAR network with a 56 Gbit/s Infiniband
network only, that limits realtime processing of the data of about five
stations. The selection of five out of the 24 stations is non-trivial as the
geometry of the stations results in very different side lobe patterns of the
final beams~\cite{Winchen2017}.

With five stations and the available processing power we expect to be able to
cover the full Moon with a grid of up to 49 analysis beams. The data-stream in
these will be constantly monitored for ns pulses during an observation and
eventually trigger a read-out of the transient buffers of all core stations to
obtain a large unfiltered data set for offline analysis.

\section{Trigger Concept}
\begin{figure}[b]
	\includegraphics[width=.49\textwidth]{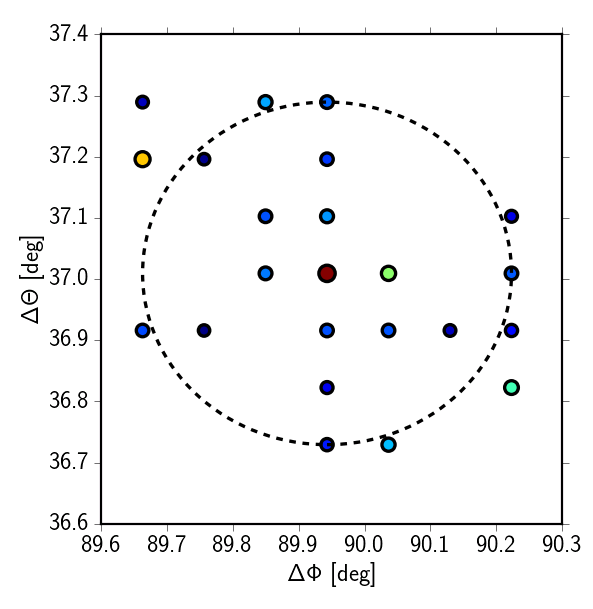}
	\includegraphics[width=.49\textwidth]{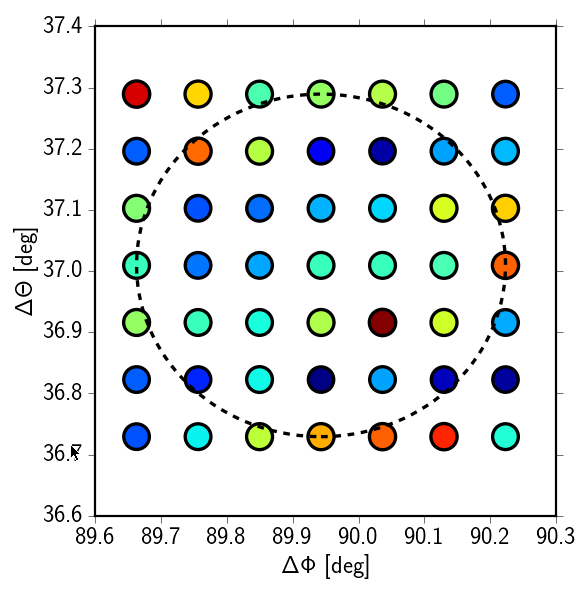}
	\caption{Amplitude above threshold of reconstructed pulses in a grid of $7\times7$ beams covering the area around the Moon. The dashed circle indicates the position and extent of the Moon. \textbf{(left)} Pulse originating from the center of the Moon. \textbf{(right)} Pulse originating from the horizon.}
	\label{fig:trigger_pattern}
\end{figure}
We expect the data to contain many RFI pulses originating from the horizon. A
background pulse originating from the horizon will produce a signal in every
analysis beam, while a signal pulse from the target direction on the sky will
be, ideally, only visible in the beam pointing to the direction of its origin.
RFI suppression can thus be, in principle, easily be achieved by requiring an
anti-coincidence between the individual analysis beams.  However, due to the circular
geometry of the beams, and also due to side-lobes the beams overlap, which
prevents a simple anti-coincidence and requires a more sophisticated trigger. An example for the reconstructed pulse amplitude for simulated
signals originating from the Moon and the horizon are shown in
figure~\ref{fig:trigger_pattern}.

We consequently implement the following prototype trigger strategy. First, the
signal in individual analysis beams is monitored and a first level trigger is
emitted when the signal passes a threshold corresponding to an acceptable
trigger rate from thermal noise of one event per minute. The number of beams
above threshold in a given time window, and the RMS of the signal amplitudes of
all beams above threshold, is then used as second criterion for signal /
background separation.

We evaluated this procedure using simulated events of varying strength
originating from the horizon and  from a random position on the Moon.  For weak
pulses that are visible in less than 10 beams we achieve a  90\% efficiency and
90\% purity using a simple cut on the normalized RMS of the amplitudes. For
strong pulses that are visible in every analysis beam we achieve 100\%
efficiency at 95\% purity. The anti-coincidence criterion thus does not limit
the method to low energy events as has been previously
remarked~\cite{Bray2016}.

\section{Expected Sensitivity}
\begin{figure}[tb]
	\includegraphics[width=\textwidth]{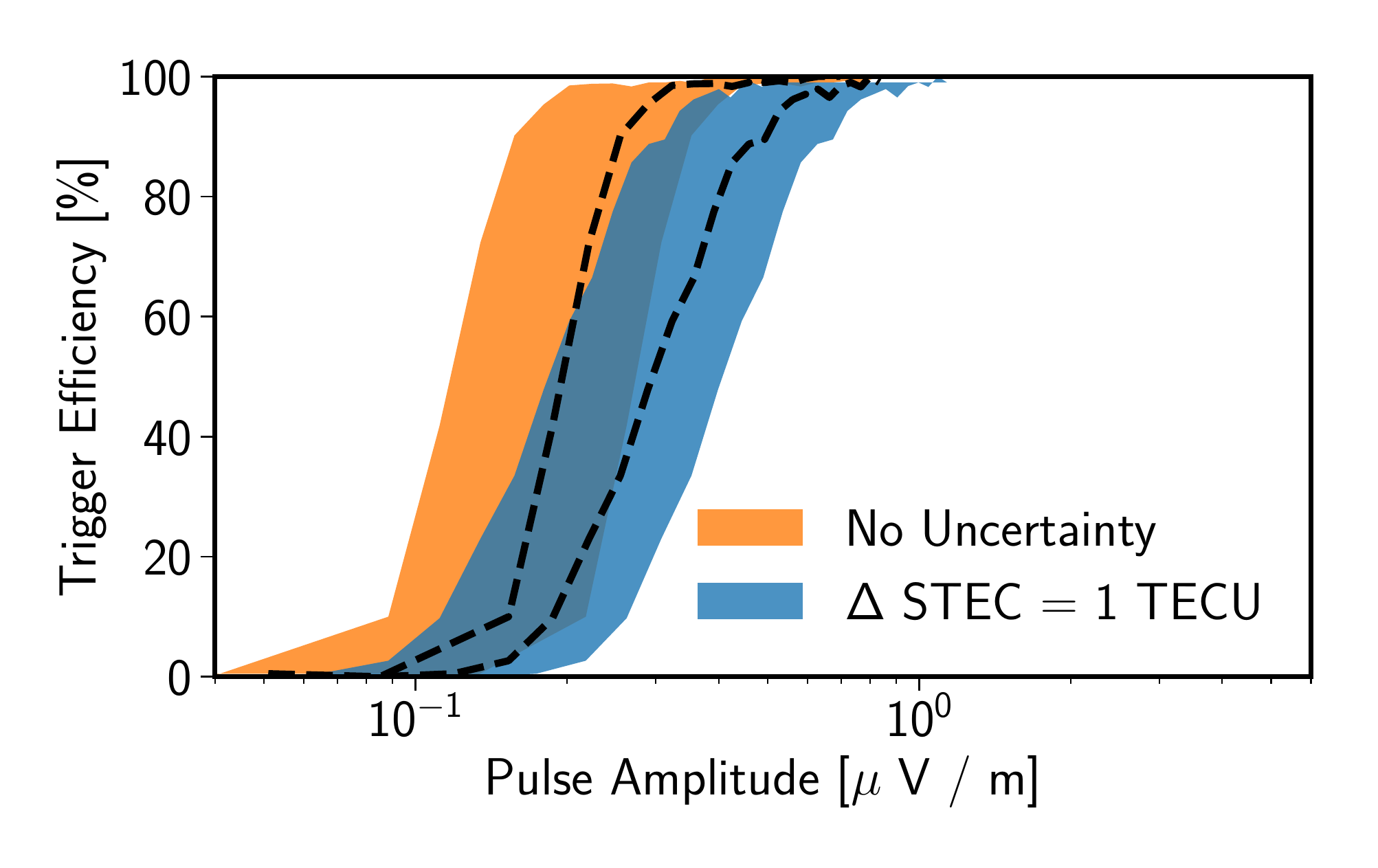}
\caption{Trigger efficiency as function of pulse amplitude for simulated
			pulses added to measured background. Pulses have been dedispersed assuming an ionospheric electron content of 20 TECU and dispersed by the same amount (orange) respectively a random value of $20 \pm 1$ TECU (blue).
 The bands denote the uncertainty on the pulse amplitude given by the uncertainty of the antenna calibration. }
	\label{fig:TriggerEfficiency}
\end{figure}
The expected sensitivity of the pipeline is estimated using simulated signal
pulses of varying strength that are added to observed noise traces.
The simulated pulses are dispersed assuming an exact slant total electron
content (STEC) of 20 TECU, respectively a normal distributed STEC value
centered around 20 TECU with width of 1 TECU to account for imperfect
reconstruction.  All traces were then processed by the observation pipeline with an
dedispersion corresponding to 20 TECU. We require one sample above a threshold value in at least one out of 49 traces for analysis beams forming a grid covering the Moon as trigger condition.
The threshold value is chosen to yield a trigger rate
of \SI{1}{\per\min} from thermal noise only. The resulting trigger efficiency is shown in
figure~\ref{fig:TriggerEfficiency} as a
function of the pulse amplitude.
Full efficiency is reached for pulses with amplitude $E \approx
\SI{0.8}{\micro\volt\per\meter}$ and approximately 20\% efficiency is reached for pulses with $E
\approx \SI{0.2}{\micro\volt\per\meter}$ if the STEC is known to 1 TECU.

\begin{figure}[tb]
	\includegraphics[width=\textwidth]{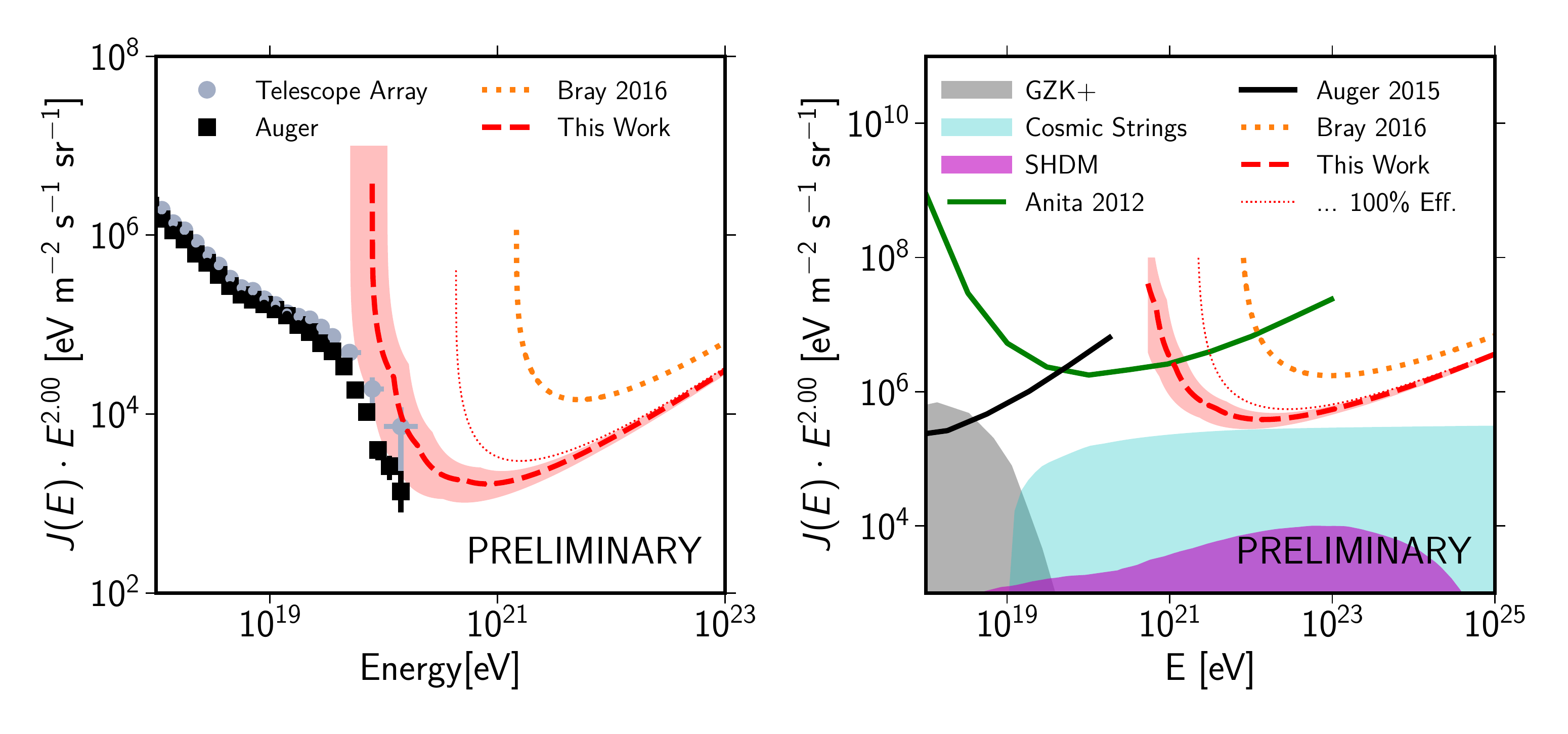}
	\caption{Expected limits for UHECR (left) and neutrinos (right) with
	available data~\cite{Fenu2017,Ivanov2012} for the UHECR flux respectively
	model predictions for the flux of neutrinos~\cite{Roulet2013,Berezinsky2011,Aloisio2015} for 200 hours of observation. The dashed line labeled 100\% Eff.\ corresponds to a minimum detectable field strength of E = \SI{0.8}{\micro\volt\per\meter}.}
	\label{fig:ExpectedLimits}
\end{figure}
Following the procedure described in references~\cite{Gayley2009, Bray2016} we
calculate the sensitivity to charged cosmic rays and neutrinos as shown in
figure~\ref{fig:ExpectedLimits}. For this calculation we assume that any reduction in observation time due to the RFI trigger rate is negligible. With the pipeline described here we expect an
improved sensitivity compared to previous results~\cite{Bray2016}, because we
can use a lower trigger threshold and an increased band-width. With the full
simulation we were also, for the first time, able to include the effect of the smooth
cut-off of sensitivity to low amplitude pulses. This softens the low-energy cut
off of the sensitivity compared to previous estimates.

\section{Conclusion}
We developed a complete prototype for the online and offline analysis pipeline
for the Lunar detection of cosmic particles with the LOFAR radio telescope.  For
the first time the sensitivities of corresponding observations have been estimated
based on full simulations of the radio telescope, data processing, and trigger. These simulations enabled design
and evaluation of new trigger concepts and consequently lead to an improved
estimated sensitivity to cosmic particles. Compared to previous semi-analytical
estimates the full detector simulations reveal an increased sensitivity to low
energetic particles of the prototype pipeline. However, with the available
resources a detection of cosmic rays with LOFAR remains unlikely, but it will
become possible by implementing the here described prototype at the upcoming
SKA telescope. Preparations for first observations to demonstrate the
capabilities of the prototype  pipeline are currently ongoing.

\section*{Acknowledgements}
The LOFAR cosmic ray key science project acknowledges funding from an Advanced Grant
of the European Research Council (FP/2007-2013) / ERC Grant Agreement n. 227610. The project
has also received funding from the European Research Council (ERC) under the European Union's
Horizon 2020 research and innovation programme (grant agreement No 640130). We furthermore
acknowledge financial support from FOM, (FOM-project 12PR3041-3) and NWO (Top Grant 614-
001-454, and Spinoza Prize SPI 78-409). TW is supported by DFG grant WI 4946/1-1.
LOFAR, the Low Frequency Array designed and constructed by ASTRON, has facilities
in several countries, that are owned by various parties (each with their own funding sources),
and that are collectively operated by the International LOFAR Telescope foundation under a joint
scientific policy.
\section*{References}

\providecommand{\newblock}{}

\end{document}